\def\grl{{GR$_\Lambda$}}
               \def\pd{\partial}
\def\dis{\displaystyle}          
\def\vsm{\vspace{-3pt}}          
\def\Leff{\hbox{$\mit\L_{\hspace{.6pt}\rm eff}$}}
\def\bull{\raise.25ex\hbox{\vrule height.8ex width.8ex}}

     \def\cM{{\cal M }}    \def\cK{{\cal K}}
\def\cO{{\cal O}}     \def\cE{{\cal E}}     \def\cH{{\cal H}}
\def\tR{{\tilde R}}   \def\tG{{\tilde G}}   \def\bL{{\bar L}}
\def\G{\Gamma}                \def\L{{\mit\Lambda}}
\def\a{\alpha}        \def\b{\beta}         
\def\d{\delta}        \def\m{\mu}           \def\n{\nu}
\def\th{\theta}               
\def\vphi{\varphi}    \def\ve{\varepsilon}  \def\Om{\Omega}
           \def\r{\rho}          \def\om{\omega}
\def\tom{{\tilde\omega}}
\def\nn{\nonumber}
\def\be{\begin{equation}}             \def\ee{\end{equation}}
\def\ba#1{\begin{array}{#1}}          \def\ea{\end{array}}
\def\bea{\begin{eqnarray} }           \def\eea{\end{eqnarray} }
\def\beann{\begin{eqnarray*} }        \def\eeann{\end{eqnarray*} }
\def\beal{\begin{eqalign}}            \def\eeal{\end{eqalign}}
\def\lab#1{\label{eq:#1}}             \def\eq#1{(\ref{eq:#1})}
\def\bsubeq{\begin{subequations}}     \def\esubeq{\end{subequations}}
\def\bitem{\begin{itemize}}           \def\eitem{\end{itemize}}

\documentclass[a4paper]{jpconf}
\usepackage{equations}

\begin{document}

\title{Asymptotic charges in 3d gravity with torsion}

\author{M Blagojevi\'c and B Cvetkovi\'c\footnote{Talk presented at
{\it Constrained Dynamics and Quantum Gravity 05\/}, Cala Gonone
(Sardinia, Italy), September 12-16, 2005.}}

\address{Institute of Physics, P.O.Box 57, 11001 Belgrade, Serbia}

\ead{mb@phy.bg.ac.yu, cbranislav@phy.bg.ac.yu}

\begin{abstract}
We discuss some new developments in three-dimensional gravity with
torsion, based on Riemann-Cartan geometry. Using the canonical
approach, we study the structure of asymptotic symmetry, clarify its
fundamental role in defining the gravitational conserved charges, and
explore the influence of the asymptotic structure on the black hole
entropy.
\end{abstract}

\section{Introduction}

Although general relativity (GR) successfully describes all the known
observational data, such fundamental issues as the nature of classical
singularities and the problem of quantization remain without answer.
Faced with such difficulties, one is naturally led to consider
technically simpler models that share the same conceptual features with
GR. A particularly useful model of this type is three-dimensional (3d)
gravity \cite{1,2,3,4,5}.

Following a widely spread belief that GR is the most reliable
approach to describe the gravitational phenomena, 3d gravity has been
studied mainly in the realm of Riemannian geometry. However, there is
a more general conception of gravity, based on {\it Riemann-Cartan
geometry\/} \cite{6}, in which both the {\it curvature\/} and the
{\it torsion\/} are used to describe the gravitational dynamics.
Here, we focus our attention on some new developments in 3d gravity,
in the realm of Riemann-Cartan geometry \cite{7,8,9,10}. We show
that the symmetry of anti-de Sitter asymptotic conditions is
described by two independent Virasoro algebras with different central
charges, in contrast to GR. We also derive the expressions for the
related conserved charges, energy and angular momentum, and discuss
the new form of the black hole entropy.

\section{Basic dynamical features}
\setcounter{equation}{0}

Theory of gravity with torsion can be formulated as Poincar\'e gauge
theory (PGT), with an underlying geometric structure described by
Riemann-Cartan space \cite{6}.

Basic gravitational variables in PGT are the triad field $b^i$ and the
Lorentz connection $A^{ij}=-A^{ji}$ (1-forms). The corresponding field
strengths are the torsion and the curvature: $T^i= db ^i+A^i{_m}\wedge
b^m$, $R^{ij}=dA^{ij}+A^i{_m}\wedge A^{mj}$ (2-forms). Gauge symmetries
of the theory are local translations and local Lorentz rotations,
parametrized by $\xi^\m$ and $\ve^{ij}$.

In 3D, we simplify the notation by introducing
$A^{ij}=-\ve^{ijk}\om_k$, $R^{ij}=-\ve^{ijk}R_k$,
$\ve^{ij}=-\ve^{ijk}\th_k$. In local coordinates $x^\m$, we have
$b^i=b^i{_\m}dx^\m$, $\om^i=\om^i{_\m}dx^\m$. The field strengths take
the form
\be
T^i= db^i+\ve^i{}_{jk}\om^j\wedge b^k\, ,\qquad
R^i=d\om^i+\frac{1}{2}\,\ve^i{}_{jk}\om^j\wedge\om^k\, ,   \lab{2.1}
\ee
and gauge transformations are
\bea
&&\d_0 b^i{_\m}= -\ve^i{}_{jk}b^j{}_{\m}\th^k-(\pd_\m\xi^\r)b^i{_\r}
     -\xi^\r\pd_\r b^i{}_\m \, ,                           \nn\\
&&\d_0\om^i{_\m}= -\nabla_\m\th^i-(\pd_\m\xi^\r)\om^i{_\r}
     -\xi^\r\pd_\r\om^i{}_\m \, ,                          \lab{2.2}
\eea
where $\nabla_\m\th^i=\pd_\m\th^i+\ve^i{}_{jk}\om^j{_\m}\th^k$ is the
covariant derivative of $\th^i$.

To clarify the geometric meaning of PGT, we introduce the metric
tensor as a bilinear combination of the triad fields:
$g=\eta_{ij}b^i\otimes b^j\equiv g_{\m\n}dx^\m\otimes dx^\n$, where
$\eta_{ij}=(+,-,-)$.
Although metric and connection are in general independent geometric
objects, in PGT they are related to each other by the {\it metricity
condition\/}: $\nabla g=0$. Consequently, the geometric structure of
PGT is described by {\it Riemann-Cartan geometry\/}.
The metricity condition implies the identity
\be
\om^i=\tom^i+K^i\, ,                                       \lab{2.3}
\ee
where $\tom^i$ is Riemannian connection,
$K_{ijk}=-\frac{1}{2}(T_{ijk}-T_{kij}+T_{jki})$ is the contortion,
and $K^i$ is defined by $K^{ij}{_m}b^m\equiv K^{ij}=-\ve^{ijk}K_k$.

General gravitational dynamics is defined by Lagrangians which are at
most quadratic in field strengths. Omitting the quadratic terms, Mielke
and Baekler proposed a {\it topological\/} model for 3D gravity
\cite{7}, defined by the action
\bea
&&I=aI_1+\L I_2+\a_3I_3+\a_4I_4+I_M\, ,                    \nn\\
&&I_1\equiv 2\int b^i\wedge R_i\, , \hspace{138pt}
  I_2\equiv-\frac{1}{3}\,\int\ve_{ijk}b^i\wedge b^j\wedge b^k\,,\nn\\
&&I_3\equiv\int\left(\om^i\wedge d\om_i
  +\frac{1}{3}\ve_{ijk}\om^i\wedge\om^j\wedge\om^k\right)\, ,
\qquad I_4\equiv\int b^i\wedge T_i\, ,                     \lab{2.4}
\eea
where $I_M$ is a matter contribution. The first term, with $a=1/16\pi
G$, is the usual Einstein-Cartan action, the second term is a
cosmological term, $I_3$ is the Chern-Simons action for the Lorentz
connection, and $I_4$ is a torsion counterpart of $I_1$. The
Mielke-Baekler model is a natural generalization of Riemannian GR with
a cosmological constant (\grl).

For isolated gravitational systems, gravitational sources can be
practically ignored in the asymptotic region. Hence, the asymptotic
structure of the theory is determined by the vacuum field equations.
In the sector $\a_3\a_4-a^2\ne 0$, these equations take the simple
form
\bea
&&2T^i=p\ve^i{}_{jk}\,b^j\wedge b^k\, ,\qquad
  p\equiv\frac{\a_3\L+\a_4 a}{\a_3\a_4-a^2} \, ,           \nn\\
&&2R^i=q\ve^i{}_{jk}\,b^j\wedge b^k\, ,
\hspace{26pt} q\equiv-\frac{(\a_4)^2+a\L}{\a_3\a_4-a^2}\, .\lab{2.5}
\eea
Thus, the vacuum configuration is characterized by constant torsion
and constant curvature. For $p=0$ or $q=0$, the vacuum geometry is
Riemannian ($T^i=0$) or teleparallel ($R^i=0$).

In Riemann-Cartan spacetime, one can use the identity (2.3) to
express the curvature $R^i(\om)$ in terms of its Riemannian piece
$\tR^i\equiv R^i(\tom)$ and the contortion:
$R^i(\om)=\tR^i+ \nabla K^i-\frac{1}{2}\ve^{imn}K_m\wedge K_n$.
This result, combined with the field equations \eq{2.5}, leads to
\be
2\tR^i=\Leff\,\ve^i{}_{jk}\,b^j\wedge b^k\, ,
 \qquad \Leff\equiv q-\frac{1}{4}p^2\, ,                   \lab{2.6}
\ee
where $\Leff$ is the effective cosmological constant. Thus, our
spacetime is maximally symmetric: for $\Leff<0$ ($\Leff\ge 0$), the
spacetime manifold is anti-de Sitter (de Sitter, Minkowski). In what
follows, our attention will be focused on the {\it anti-de Sitter\/}
sector: $\Leff=-1/\ell^2<0$.

\section{The black hole with torsion}
\setcounter{equation}{0}

For $\Leff<0$, equation \eq{2.6} has a well known solution for the
metric --- the BTZ black hole. In the static coordinates
$x^\m=(t,r,\vphi)$ with $0\le\vphi<2\pi$, the black hole metric is
given by
\bea
&&ds^2=N^2dt^2-N^{-2}dr^2-r^2(d\vphi+N_\vphi dt)^2\, ,     \nn\\
&&N^2=\left(-8Gm+\frac{r^2}{\ell^2}+\frac{16G^2J^2}{r^2}\right)\, ,
  \qquad N_\vphi=\frac{4GJ}{r^2}\, ,                       \lab{3.1}
\eea
Using local Lorentz invariance, we can choose $b^i$ to have the
simple, ``diagonal" form:
\bsubeq\lab{3.2}
\be
b^0=Ndt\, ,\qquad b^1=N^{-1}dr\, ,\qquad
b^2=r\left(d\vphi+N_\vphi dt\right)\, .                    \lab{3.2a}
\ee
To find the connection, we combine the relation $K^i=(p/2)b^i$, which
follows from the first field equation in \eq{2.5}, with the identity
\eq{2.3}. This yields
\bea
&&\om^i=\tom^i+\frac{p}{2}\,b^i\, ,                        \nn\\
&&\tom^0=-Nd\vphi\, ,\qquad \tom^1=N^{-1}N_\vphi dr\, ,\qquad
  \tom^2=-\frac{r}{\ell^2}dt-rN_\vphi d\vphi\, .           \lab{3.2b}
\eea
\esubeq
where $\tom^i$ is Riemannian connection, defined by
$d\tom^i+\ve^i{}_{jk}\tom^j b^k=0$. Equations \eq{3.2} define the
analogue of the BTZ black hole in {\it Riemann--Cartan\/} spacetime
\cite{8,9}.

As a constant curvature spacetime, the black hole is locally isometric
to the AdS solution (AdS$_3$), obtained formally from \eq{3.2} by the
replacement $J=0$, $8Gm=-1$.

\section{Asymptotic conditions}
\setcounter{equation}{0}

For isolated gravitational systems, matter is absent from the
asymptotic region, but it can influence global properties of
spacetime through the asymptotic conditions. The symmetries of these
conditions are closely related to the gravitational conserved charges
\cite{10}.

For $\Leff<0$, maximally symmetric AdS solution has the role
analogous to the role of Minkowski space in the $\Leff=0$ case.
Following the analogy, we could choose that all the fields approach
the single AdS$_3$ configuration at large distances, leading to the
global AdS symmetry. However, this choice would exclude the important
black hole solution. This motivates us to introduce the {\it
asymptotic AdS configurations\/}, determined by the following
requirements:
\bitem
\item[(a)] the asymptotic conditions include the black hole
configuration,  \vsm
\item[(b)] they are invariant under the action of the AdS group
$SO(2,2)$, and  \vsm
\item[(c)] the asymptotic symmetries have well defined canonical
generators.
\eitem

The asymptotics of the triad field $b^i{_\m}$ that satisfies (a) and
(b) reads:
\bsubeq\lab{4.1}
\be
b^i{_\m}= \left( \ba{ccc}
   \dis\frac{r}{\ell}+\cO_1   & O_4  & O_1     \\
   \cO_2 & \dis\frac{\ell}{r}+\cO_3  & O_2     \\
   \cO_1 & \cO_4                     & r+\cO_1
                 \ea
          \right) \, .                                     \lab{4.1a}
\ee
Here, for any $\cO_n=c/r^n$, we assume that $c$ is not a constant,
but a function of $t$ and $\vphi$, $c=c(t,\vphi)$, which is the
simplest way to ensure the global $SO(2,2)$ invariance.

The asymptotic form of $\om^i{_\m}$ is defined in accordance with
\eq{3.2b}:
\be
\om^i{_\m}=\left( \ba{ccc}
   \dis\frac{pr}{2\ell}+\cO_1 & \cO_4 &\dis-\frac{r}{\ell}+\cO_1\\
   \cO_2 & \dis\frac{p\ell}{2r}+\cO_3 & \cO_2                   \\
   \dis-\frac{r}{\ell^2}+\cO_1 & \cO_4 & \dis\frac{pr}{2}+\cO_1
                  \ea
           \right) \, .                                    \lab{4.1b}
\ee
\esubeq
A verification of the third condition (c) is left for the next section.

Having chosen the asymptotic conditions, we now wish to find the subset
of gauge transformations \eq{2.2} that respect these conditions. They
are defined by restricting the original gauge parameters in accordance
with \eq{4.1}, which yields
\bea
&&\xi^0=\ell\left[
  T+\frac{1}{2}\left(\frac{\pd^2 T}{\pd t^2}\right)
                     \frac{\ell^4}{r^2}\right]+\cO_4\, ,\quad
  \xi^1=-\ell\left(\frac{\pd T}{\pd t}\right)r+\cO_1\, ,\quad
  \xi^2=S-\frac{1}{2}\left(\frac{\pd^2 S}{\pd\vphi^2}\right)
                     \frac{\ell^2}{r^2}+\cO_4\, ,          \nn\\
&&\th^0=-\frac{\ell^2}{r}\pd_0\pd_2T+\cO_3\, ,\quad
  \th^1=\pd_2 T+\cO_2\, ,\quad
  \th^2=\frac{\ell^3}{r}\pd_0^2T+\cO_3\, .                 \lab{4.2}
\eea
The functions $T$ and $S$ are such that $\pd_\pm(T\mp S)=0$, with
$x^\pm\equiv x^0/\ell \pm x^2$, which implies $T+S=g(x^+)$,
$T-S=h(x^-)$, where $g$ and $h$ are two arbitrary, periodic functions.

The commutator algebra of the Poincar\'e gauge transformations
\eq{2.2} is closed: $[\d_0',\d_0'']=\d_0'''$, where
$\d_0'=\d_0(\xi',\th')$ and so on. Using the related composition law
with the restricted parameters \eq{4.2}, and keeping only the lowest
order terms, one finds the relation
\bea
&&T'''=T'\pd_2 S''+S'\pd_2 T''-T''\pd_2 S'-S''\pd_2 T'\, , \nn\\
&&S'''=S'\pd_2 S''+T'\pd_2 T''-S''\pd_2 T'-T''\pd_2 ST'\, .\lab{4.3}
\eea
Let us separate the parameters \eq{4.2} into two pieces: the leading
terms containing $T$ and $S$ define a $(T,S)$ transformation, while
the rest defines the residual (pure gauge) transformation. The PGT
commutator algebra implies that the commutator of two $(T,S)$
transformations produces not only a $(T,S)$ transformations, but also
an additional pure gauge transformation. This result motivates us to
introduce an improved definition of the asymptotic symmetry: it is
the symmetry defined by the parameters \eq{4.2}, modulo pure gauge
transformations. As we shall see in the next section, this symmetry
coincides with the {\it conformal symmetry\/}.

\section{Canonical generators and conserved charges}
\setcounter{equation}{0}

We continue our study of the asymptotic symmetries and conservation
laws in the canonical formalism \cite{10}. Introducing the canonical
momenta $(\pi_i{^\m},\Pi_i{^\m})$, corres\-pon\-ding to the Lagrangian
variables $(b^i{_\m},\om^i{_\m})$, we find that the primary constraints
of the theory \eq{2.4} are of the form:
\bea
&&\phi_i{^0}\equiv\pi_i{^0}\approx 0\, ,\hspace{86.5pt}
   \Phi_i{^0}\equiv\Pi_i{^0}\approx 0\, ,                  \nn\\
&&\phi_i{^\a}\equiv\pi_i{^\a}
  -\a_4\ve^{0\a\b}b_{i\b}\approx 0\, ,\qquad
   \Phi_i{^\a}\equiv\Pi_i{^\a}
  -\ve^{0\a\b}(2ab_{i\b}+\a_3\om_{i\b})\approx 0\, .       \nn
\eea
Up to an irrelevant divergence, the total Hamiltonian reads
\bea
&&\cH_T=b^i{_0}\cH_i+\om^i{_0}\cK_i
   +u^i{_0}\pi_i{^0}+v^i{_0}\Pi_i{^0} \, ,                 \nn\\
&&\cH_i=-\ve^{0\a\b}\left(aR_{i\a\b}+\a_4T_{i\a\b}
   -\L \ve_{ijk}b^j{_\a}b^k{_\b}\right)-\nabla_\b\phi_i{^\b}
   +\ve_{imn}b^m{_\b}\left(p\phi^{n\b}+q\Phi^{n\b}\right)\,,\nn\\
&&\cK_i=-\ve^{0\a\b}\left(aT_{i\a\b}+\a_3R_{i\a\b}
   +\a_4\ve_{imn}b^m{_\a}b^n{_\b}\right)
   -\nabla_\b\Phi_i{^\b}-\ve_{imn}b^m{_\b}\phi^{n\b}\, .   \nn
\eea
The constraints ($\pi_i{^0},\Pi_i{^0},\cH_i,\cK_i$) are first class,
($\phi_i{^\a},\Phi_i{^\a}$) are second class.

Applying the general Castellani's algorithm \cite{6}, we find the
canonical gauge generator:
\bea
&& G=-G_1-G_2\, ,                                          \lab{5.1}\\
&&G_1\equiv\dot\xi^\r\left(b^i{_\r}\pi_i{^0}+\om^i{_\r}\Pi_i{^0}\right)
  +\xi^\r\left[b^i{_\r}\cH_i +\om^i{_\r}\cK_i
  +(\pd_\r b^i{_0})\pi_i{^0}+(\pd_\r\om^i{_0})\Pi_i{^0}\right]\,,\nn\\
&&G_2\equiv\dot\th^i\Pi_i{^0}
  +\th^i\left[\cK_i-\ve_{ijk}\left( b^j{_0}\pi^{k0}
  +\om^j{_0}\,\Pi^{k0}\right)\right]\, .                   \nn
\eea
Here, the time derivatives $\dot b^i{_\m}$ and $\dot\om^i{_\m}$ are
shorts for $u^i{_\m}$ and $v^i{_\m}$, respectively, and the integration
symbol $\int d^2x$ is omitted in order to simplify the notation. The
transformation law of the fields, $\d_0\phi\equiv \{\phi\,,G\}$, is in
complete agreement with the gauge transformations \eq{2.2} {\it on
shell\/}.

The behaviour of momentum variables at large distances is defined by
the following general principle: the expressions that vanish on-shell
should have an arbitrarily fast asymptotic decrease, as no solution of
the field equations is thereby lost.

The canonical generator acts on dynamical variables via the Poisson
bracket operation, which is defined in terms of functional
derivatives. In general, $G$ does not have well defined functional
derivatives, but the problem can be corrected by adding suitable {\it
surface terms\/} \cite{6}. The improved canonical generator $\tG$
reads:
\bea
&&\tG=G+\G\, ,\qquad
  \G\equiv-\int_0^{2\pi}d\vphi
         \left(\xi^0\cE^1+\xi^2\cM^1\right)\, ,            \lab{5.2}\\
&&\cE^\a\equiv
    2\ve^{0\a\b}\left[\left(a+\frac{\a_3p}{2}\right)\om^0{}_\b
   +\left(\a_4+\frac{ap}{2}\right)b^0{}_\b+\frac{a}{\ell}b^2{}_\b
   +\frac{\a_3}{\ell}\om^2{}_\b\right]b^0{}_0\, ,          \nn\\
&&\cM^\a\equiv
   -2\ve^{0\a\b}\left[\left(a+\frac{\a_3p}{2}\right)\om^2{}_\b
   +\left(\a_4+\frac{ap}{2}\right)b^2{}_\b+\frac{a}{\ell}b^0{}_\b
   +\frac{\a_3}{\ell}\om^0{}_\b\right]b^2{}_2\, .          \nn
\eea
The adopted asymptotic conditions guarantee differentiability and
finiteness of $\tG$. Moreover, $\tG$ is also {\it conserved\/}.

The value of the improved generator $\tG$ defines the {\it
gravitational charge\/}. Since $\tG\approx \G$, the charge is
completely determined by the boundary term $\G$. Note that $\G$ depends
on $T$ and $S$, but not on pure gauge parameters. For $\xi^2=0$, $\tG$
reduces to the time translation generator, while for $\xi^0=0$ we
obtain the spatial rotation generator. The corresponding surface terms,
calculated for $\xi^0=1$ and $\xi^2=1$, respectively, have the meaning
of {\it energy\/} and {\it angular momentum\/}:
\be
E=\int_0^{2\pi}d\vphi\,\cE^1 \, ,\qquad
M=\int_0^{2\pi}d\vphi\,\cM^1 \, .                          \lab{5.3}
\ee
Energy and angular momentum are conserved gravitational charges.

Using these results, one can calculate the conserved charges for the
black hole \cite{8,9}:
\be
E= m+\frac{\a_3}{a}\left(\frac{pm}{2}-\frac{J}{\ell^2}\right)\, ,
\qquad M= J+\frac{\a_3}{a}\left(\frac{pJ}{2}-m\right)\, .  \lab{5.4}
\ee
They differ from the corresponding expressions in Riemannian \grl,
where $\a_3=0$.

\section{Canonical algebra}
\setcounter{equation}{0}

The structure of the asymptotic symmetry is encoded in the Poisson
bracket algebra of the improved generators. In the notation $G'\equiv
G[T',S']$, $G''\equiv G[T'',S'']$, and so on, the Poisson bracket
algebra is found to have the form $\left\{\tG'',\,\tG'\right\}
=\tG''' + C'''$, where the parameters $T'''$, $S'''$  are determined
by the composition rules  \eq{4.3}, and $C'''$ is the {\it central
term\/} of the algebra. Expressed in terms of the Fourier modes, this
algebra takes a more familiar form---the form of two independent
Virasoro algebras with classical central charges:
\bea
&&\left\{L_n,L_m\right\}
   =-i(n-m)L_{n+m}-\frac{c}{12}in^3\d_{n,-m}\, ,           \nn\\
&&\left\{\bL_n,\bL_m\right\}
   =-i(n-m)\bL_{m+n}-\frac{\bar c}{12}in^3\d_{n,-m}\, ,    \lab{6.1}
\eea
and $\{L_n,\bar{L}_m\}=0$. The central charges have the form:
\be
c=\frac{3\ell}{2G}+24\pi\a_3\left(\frac{p\ell}{2}+1\right)\, ,\qquad
\bar{c}=\frac{3\ell}{2G}
        +24\pi\a_3\left(\frac{p\ell}{2}-1\right)\, .       \lab{6.2}
\ee
Asymptotically, the gravitational dynamics is characterized by the
conformal symmetry with two {\it different\/} central charges, in
contrast to Riemannian \grl, where $c=\bar c=3\ell/2G$.

\section{The black hole entropy}
\setcounter{equation}{0}

A particularly interesting consequence of the theory defined by the
action \eq{2.4} is that the entropy of the black hole with torsion
differs from the corresponding Riemannian result. Indeed, the
semi-classical calculation based on the Hamiltonian form of the
action leads to
\be
S=\frac{2\pi r_+}{4G}
  +4\pi^2\a_3\left(pr_+ -2\frac{r_-}{\ell}\right)\, ,      \lab{7.1}
\ee
where $r_\pm$ are the zeros of $N^2$. For \grl\ with Riemannian
Chern-Simons term ($p=0$, $\a_4=0$, but $\a_3\ne 0$), our formula
for $S$ yields Solodukhin's result \cite{11}. Using the new
expressions \eq{5.4} for energy and angular momentum, one can easily
verify that the first law of black hole thermodynamics takes the form
\be
dE=TdS+\Om dM\, ,\qquad \Om\equiv N_\vphi(r_+)\, .         \lab{7.2}
\ee
Thus, the existence of torsion is in complete agreement with the first
law of thermodynamics.\footnote{Thermodynamics of black holes is
properly formulated in the Euclidean sector.}

\section{Concluding remarks}

\indent

$\bull~$ 3d gravity with torsion, defined by the action \eq{2.4},
is based on an underlying Riemann-Cartan geometry of spacetime.

$\bull~$ The theory possesses the black hole solution \eq{3.2}, a
generalization of the Riemannian BTZ black hole. Energy and angular
momentum of the black hole differ from the corresponding Riemannian
expressions in \grl.

$\bull~$ The AdS asymptotic conditions \eq{4.1} imply the conformal
symmetry in the asymptotic region, which is described by two
independent Virasoro algebras with different central charges.

$\bull~$ The existence of different central charges ($\a_3\ne 0$)
modifies the black hole entropy, but remains in agreement with the
first law of thermodynamics.

\ack

This work was partially supported by the Serbian Science Foundation,
Serbia.

\end{document}